\documentstyle[12pt,epsfig]{article}
\title{Upper limit on the $\eta \to \pi^0 \pi^0$ decay}
\author{ M.N.~Achasov, A.V.~Berdyugin,
A.V.~Bozhenok, D.A.~Bukin,\\ S.V.~Burdin,
T.V.~Dimova, S.I.~Dolinskii, V.P.~Druzhinin, \\ M.S.~Dubrovin,  
I.A.~Gaponenko, V.B.~Golubev, V.N.~Ivanchenko, \\
A.A.~Korol, S.V.~Koshuba, I.N.~Nesterenko,
E.V.~Pakhtusova, \\ E.A.~Perevedentsev,  A.A.~Salnikov, 
S.I.~Serednyakov, V.V.~Shary, \\ Yu.M.~Shatunov, V.A.~Sidorov,
Z.K.~Silagadze\thanks {Corresponding author. Fax +7 3832 35 21 63,
e-mail silagadze@inp.nsk.su} , Yu.S.~Velikzhanin.
\vspace*{3mm} \\ Budker Institute of Nuclear Physics \\ 630 090,
Novosibirsk, Russia }
\date{}

\begin{document}
\large
\maketitle

\begin{abstract}
Upper bound ${\rm Br}( \eta \to \pi^0 \pi^0 )  < 6 \cdot
10^{-4}$ with 90 \% confidence is reported from the SND experiment 
at the VEPP-2M collider.
\end{abstract}

CP-noninvariance still remains the most mysterious phenomenon
in elementary particle physics because it implies either 
an absolute difference between left and right in nature, or maybe
even more exotic possibility of the mirror world existence \cite{1}.

CP-violation is incorporated into the Standard Model by means of 
complex coupling constants in the quark mixing matrix and is controlled by 
a single parameter - the Kobayashi-Maskawa phase \cite{2}.
The only manifestations of the CP violating interactions, observed by
now, are restricted to the $K_L$ meson decays. More profound 
CP-violating effects are predicted in the $B^0$-meson decays but the 
corresponding experiments are still awaiting their time.

Flavor-conserving CP-violating interactions, like $\eta \to 2\pi$
decay, are expected to be extremely small in the Standard Model, namely
${\rm Br}( \eta \to \pi^+ \pi^- )  \le 2 \cdot 10^{-27} $ \cite{3,4}.
Nevertheless, experimental search for the  $\eta \to 2\pi$ decay
is very important \cite{5}, because unconventional CP-violating effects
have "a golden opportunity" to reveal themselves in this decay. 

Let us mention two of such CP-violation models beyond the Standard 
Model. $\eta \to 2\pi$ decay can take place if there exists a strong
CP-violation due to QCD $\Theta$-term \cite{6}. Using present 
experimental constraint on this $\Theta$-term from the electric dipole
moment of the neutron, one can obtain ${\rm Br}( \eta \to \pi^+ \pi^- )  
\le 0.78 \cdot 10^{-16}$ \cite{3,4}. Another possibility is a spontaneous 
CP violation in the extended Higgs sector \cite{7}. It was shown in \cite{4}
that in this case  ${\rm Br}( \eta \to \pi^+ \pi^- ) $ 
can be as high as $ 1.2 \cdot 10^{-15}$. 
 But even this last value is
too small for  $\eta \to 2\pi$ decay to be observed in any present
or planned experiments. So its observation will be a clear signal
of some CP-violation exotica.

The only experimental upper limit ${\rm Br}( \eta \to \pi^+ \pi^- )  
\le 1.5 \cdot 10^{-3} $ was reported in \cite{8}. Recently this result
was confirmed by CMD-2 detector \cite{9}. Here we present, for the
first time, upper bound ${\rm Br}( \eta \to \pi^0 \pi^0 )         
< 6 \cdot 10^{-4} $ obtained in  SND experiment at VEPP-2M $e^+e^-$ 
collider.

SND \cite{10} is a general purpose nonmagnetic detector. Its main part, very
important for neutral decay channels, is a highly segmented three layer 
NaI(Tl) calorimeter \cite{11} of 3.6 t total weight with vacuum phototriode
\cite{12} readout.

During 1996 experimental run, SND collected data corresponding to integrated
luminosity of about 3.9 ${\rm pb}^{-1}$.
The data sample consists of six successive scans of the narrow
energy interval covering the $\phi$-meson peak (FI96 experiment \cite{13}).
Total number of produced $\phi$-mesons is $7.4 \cdot 10^6$, corresponding
to $9.3 \cdot 10^4$ $\eta$-mesons, produced through $\phi \to \eta 
\gamma$ decay. Five-photon events from this data sample,  
with five clusters in the calorimeter and no charged particles,
were used to search for a signal from the $\eta \to 2\pi^0$ decay, in the 
reaction $\phi \to \eta \gamma , \; \eta \to 2\pi^0$. Only the events
having the normalized full momentum less than 0.15, and the normalized 
total energy deposition in the calorimeter greater than 0.8 were selected
for the analysis.

For events which had two $\pi^0$ candidates with
$$(m^{(1)}_{2\gamma}-m_{\pi^0})^2+(m^{(2)}_{2\gamma}-m_{\pi^0})^2 < 300
{\rm MeV}^2$$
\noindent a kinematical fit was performed assuming
energy-momentum conservation and presence of two $\pi^0$ mesons in the final 
state. $\chi^2$ of this kinematical fit can be used to reject
background from the $\phi \to K_S K_L$ decay, because these background 
events have a poor energy-momentum balance. In the present analysis we used
mild cut $\chi^2 < 40$ .

Another background comes from the $\phi \to \eta \gamma, \; \eta \to 3\pi^0$
decay. Examining the transverse energy profile of the electromagnetic shower 
in the calorimeter, one can recognize and reject events where close photons 
merge together \cite{14}. 
This merging is one of the main reasons why $3\pi^0 \gamma$ final state can
mimic $2\pi^0 \gamma$ final state. We demand the corresponding separation
parameter (described in detail in \cite{14}) to be less than 5, which is 
also a mild cut. 

Remaining events, surviving these cuts, are mainly due to $ e^+e^- \to
\omega \pi^0$, $\omega \to \pi^0 \gamma$ reaction. To suppress this
background, additional cuts $m_{\pi^0 \gamma} < 730{\rm MeV}$ and
$\chi_\omega^2>20$ were applied. Here $\chi_\omega^2$ is a parameter
describing degree of likelihood of the assumption that
our 5-photon event is a $ e^+e^- \to \omega \pi^0 \to \pi^0 \pi^0 \gamma$
event.

The photon  recoil mass distribution for remaining 70 events is shown
in fig.1. Note, that Monte-Carlo simulation predicts that $11 \pm 3$ 
background events from the $e^+e^- \to \omega \pi^0$ and
$19 \pm 5$ events from the $e^+e^-\to \eta \gamma \to 3\pi^0\gamma, \;
3\gamma, \; \dots$ reactions surviving these cuts. It is difficult to
estimate the remaining background from the $\phi \to K_S K_L$ channel
because of huge Monte-Carlo statistics and high accuracy for the $K$-mesons 
nuclear interactions simulation needed for this goal. No 
candidate events passed analysis cuts from the simulated $4 \cdot 10^5$
$\phi \to K_S K_L$ decays. 


It turns out that the most significant background comes from the
$\phi \to f^0 \gamma,\; f^0\to 2\pi^0$ channel: $52\pm 13$ events are 
expected according to MC, if the recently published parameters of
this decay \cite{15} are used in simulation.

Background from the $\phi \to \eta \gamma,\; \eta\to \pi^0 \gamma\gamma$
decay seems to be negligible. Only $0.7 \pm 0.2$ events are expected due to 
very small $\eta\to \pi^0 \gamma \gamma$ branching ratio.

In total we expect $83\pm 14$ background events from the processes 
described above, which is in a good agreement with the 
experimentally observed number. Although, in the $\pm 3\sigma$ interval 
around the expected $\eta$-peak only one experimental event was observed,
while Monte-Carlo predicts $10\pm 5$ events. Due to small efficiences for
background processes in the signal area, significant
systematic errors could be expected in their simulation. So we don't perform
any background subtraction and our result is based only on the experimental 
statistics.

Fig.1 shows that no significant signal from the $\eta \to 2\pi^0$ could be
seen over the expected background from the $f^0 \gamma$, $\omega \pi^0$
and $\eta \to 3\pi^0$ channels. Taking $\epsilon=7.5\%$ as a detection 
efficiency for
the $\eta \to 2\pi^0$ decay, which follows from the Monte-Carlo analysis,
and assuming $N_x=3.89$ for Poisson upper limit, which corresponds to the 
90 \% confidence level for one observed event \cite{16}, we get the 
upper bound
$$ {\rm Br}(\eta \to 2\pi^0) < \frac{N_x}{N_\eta \cdot \epsilon}
\approx 6 \cdot 10^{-4} \; \; . $$

The work is partially supported by RFBR (Grants No 96-02-19122, 97-02-18563).

\newpage
\begin{figure}[htb]
     \mbox{\epsfig{figure=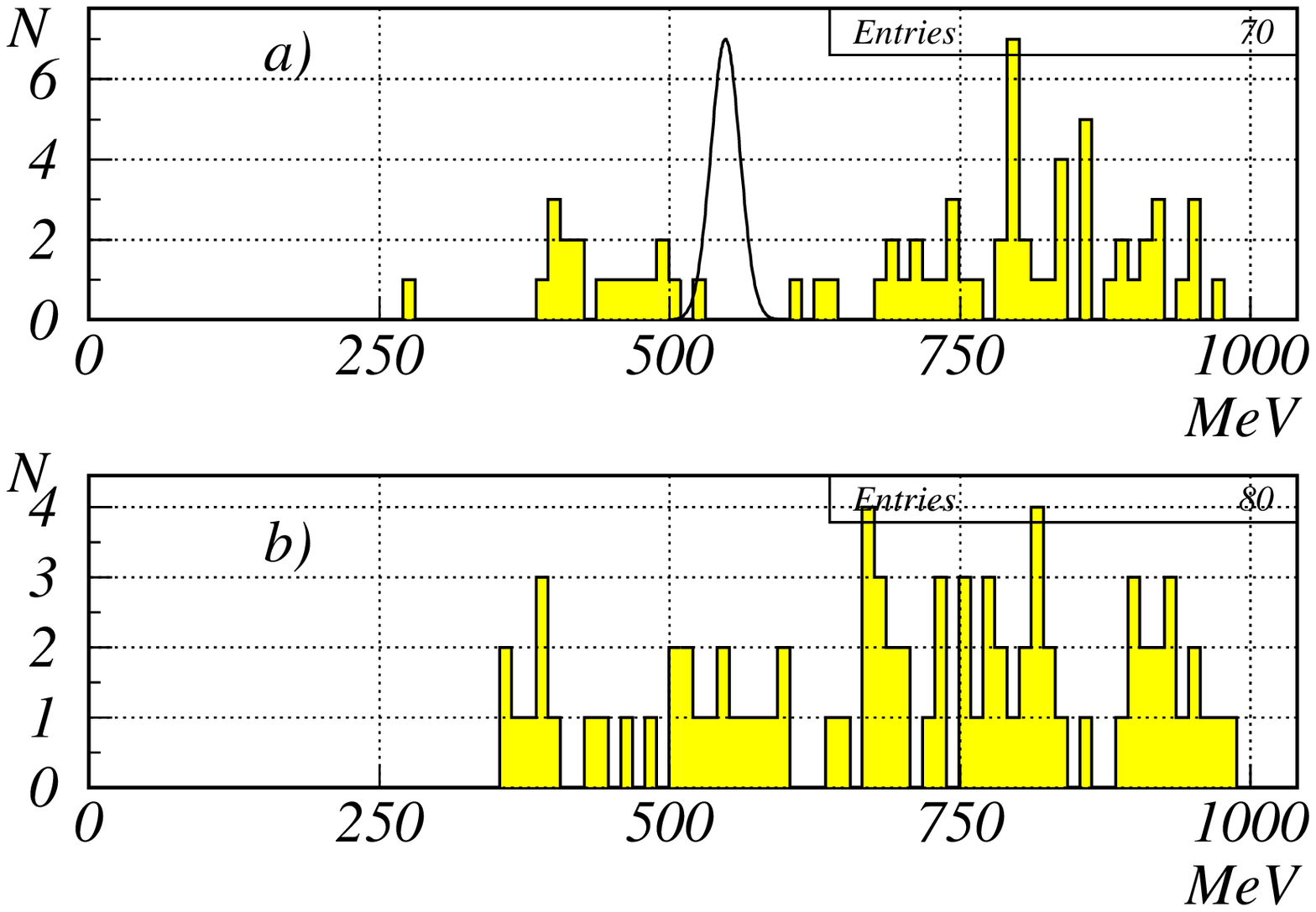 
	                       ,  height=15.0cm}}
     \caption {Photon recoil mass. a) - experiment. Solid curve indicates
     the position and the width (according to MC simulation) of a possible
     signal from the $\eta \to 2\pi^0$ decay. The peak amplitude corresponds
     to the $3 \cdot 10^{-3}$ branching ratio. 
     b) - expected background according to MC simulation.}
\label{Fig1}
\end{figure}

\end{document}